\documentstyle[epsf]{article}
\newcommand{\ndt}{\noindent}
\begin{document}
%\begin{flushright}
%{DFUB 2005/3}\\
%DRAFT, \today
%\end{flushright}
\vspace{5mm}

\begin{center}
\large
\bf{SEARCH FOR SPONTANEOUS MUON EMISSION FROM LEAD NUCLEI}\\
\vspace{7mm}
\end{center}

\ndt L. Arrabito$^5$, D. Autiero$^5$, E. Barbuto$^9$, C. Bozza$^9$, S.
Cecchini$^{2,a}$, L. Consiglio$^2$, M. Cozzi$^2$, N. D'Ambrosio$^6$,
Y. Declais$^5$, G. De Lellis$^6$, G. De Rosa$^6$, M. De Serio$^1$, D. Di
Ferdinando$^2$, A. Di Giovanni$^4$, N. Di Marco$^4$, L.S. Esposito$^2$, G.
Giacomelli$^2$, M. Giorgini$^2$, G. Grella$^9$, M. Hauger$^7$, M. Ieva$^1$,
D.B.Ion$^{3}$,
 I. Janicsko$^7$, F. Juget$^7$, I. Laktineh$^5$, G. Mandrioli$^2$,
S. Manzoor$^{2,b}$, A. Margiotta$^2$, P. Migliozzi$^6$, P.
Monacelli$^4$, M.T. Muciaccia$^1$, L. Patrizii$^2$, C. Pistillo$^6$, V.
Popa$^{2,c}$, G. Romano$^9$, G. Rosa$^8$, P. Royole-Degieux$^5$,
S. Simone$^1$, M. Sioli$^2$, C. Sirignano$^9$, G. Sirri$^2$, G.
Sorrentino$^6$, M. Spurio$^2$ and V. Tioukov$^6$

\vspace{5mm}

{\small \ndt
$^1$ Dept. of Physics, University of Bari and INFN, I-70126 Bari, Italy\\
$^2$ Dept. of Physics, University of Bologna and INFN, I-40127 Bologna, Italy\\
$^3$ National Institute for Nuclear Physics and Engineering. R-77125 Bucharest, Romania\\
$^4$ Dept. of Physics, University of L'Aquila and INFN, I-67010 L'Aquila,
 Italy \\
$^5$ IPN Lyon, F-69622 Villeurbanne, Lyon, France \\
$^6$ Dept. of Physics, University of Napoli and INFN, I-80125 Napoli, Italy\\
$^7$ Inst. de Physique, CH-2000 Neuchatel, Switzerland \\
$^8$ Dept. of Physics, University of Roma ``La Sapienza'' and INFN, I-00185
     Roma, Italy \\
$^9$ Dept. of Physics, University of Salerno and INFN, I-84081
Salerno, Italy \\
$^a$ also IASF/CNR, Bologna, Italy \\
$^b$ also PRD, PINSTECH, P.O. Nilore, Islamabad, Pakistan \\
$^c$ also Institute for Space Sciences, R-77125 Bucharest, Romania}
% I.N.F.N., Laboratori Nazionali Frascati, I-00044, Frascati (Rome), Italy

\begin{abstract}

We describe a possible search for muonic radioactivity from lead nuclei
using the base elements (``bricks'' composed by lead and nuclear emulsion
sheets) of the long-baseline OPERA neutrino experiment. We
present the results of a Monte Carlo simulation
concerning the expected event topologies and estimates of the
background events. Using few bricks, we could reach a good
sensitivity level.

\end{abstract}

\section{Introduction}
\label{sec:intro}
In the late 80's, some theoretical work was dedicated to investigate possible
exotic types of nuclear radioactivity, consisting in the emission of
light particles such
as pions or muons from heavy nuclei \cite{ion1, ion2}. Although the
theoretical estimates indicated a larger branching ratio for spontaneous
(or neutron-induced) muon emission than for pions\footnote{This observation
is not necessarly valid in the case of Pb, as the predicted branching
ratios for the pionic and muonic radioactivities are defined in \cite{ion1,ion2}
relatively to the spontaneous fission of the parent nuclei.}, the experimental
searches were mostly dedicated to the pionic radioactivity
\cite{bucurescu, rus}, since the detection conditions were more favourable
with the used experimental set-ups.
 None of the experiments using heavy radioactive
nuclei could  prove the existence of the pionic or muonic radioactivities,
 but some of them may have produced indications in support of these
hypotheses \cite{results}. Some experimental indications for pionic
radioactivity come from the interpretation of radioactive super giant
halos in crystals \cite{halo2}; such a technique, however, cannot give
precise information concerning the nature of the source or allow quantitative
estimates of branching ratios, lifetimes, etc. A short review of the
theoretical problems and of the experimental results obtained, mostly
for the pionic radioactivity, may be found in ref. \cite{ion3}.

 Muons or pions could be emitted by nuclei through the decays \cite{ion2}:
\begin{equation}
(A,Z) \rightarrow \mu^\pm + \nu_\mu(\overline \nu_\mu)+(A_1,Z_1)+...+(A_n,Z_n),
\label{eq:1}
\end{equation}
\begin{equation}
(A,Z) \rightarrow \pi^\pm (\pi^0)+(A_1,Z_1)+...+(A_n,Z_n),
\label{eq:2}
\end{equation}
where, for reasons of energy and momentum conservation, the number
of fragments $n$ is $\geq 2$. The nuclei $(A_1,Z_1),~(A_2,Z_2),...,(A_n,Z_n)$ 
would yield a sequence of $\beta^-$ decays leading finally to stable nuclei 
with a balanced neutron-proton ratio.

In ref. \cite{ion2} some nuclear charge thresholds 
for different possible spontaneous particle emission were listed: \par

 $(i)~ \mu^\pm$ (prompt muon) for $Z \geq 72$ \par
 $(ii)~ \pi^\pm$ (prompt pion $\rightarrow$ delayed muon) for $Z \geq 76$ \par 
 $(iii)~ 2\mu^\pm$ (prompt muon pairs) for $Z \geq 91$ \par
 $(iv)~2\pi^\pm$ (prompt pion pairs $\rightarrow$  delayed muon pairs) 
for $Z \geq 100$ 

Natural lead is mainly composed by three nuclides: $^{206}$Pb (24.1\%), 
 $^{207}$Pb (22.1\%) and $^{208}$Pb (52.4\%). They are stable nuclides, but 
the channels $(i)$ and $(ii)$ are energetically allowed. 

The spontaneous or neutron induced fission of Pb has never been observed. 
Nevertheless, the decay into a lepton-antilepton pair (muon and neutrino) 
or into a quark-antiquark pair (forming a pion) could enhance the probability
that the remaining ``hyper-cold'' nuclear state be unstable \cite{ion1}. 

A search for a Pb muonic decay can be made as a byproduct of the OPERA 
experiment \cite{opera}. Uranium or Thorium could be used in the future; such
possible experiments would benefit from the results of an initial search 
using Lead.  

In the hypothesis of the decays (\ref{eq:1}) and (\ref{eq:2}), the 
fission fragments would remain nearly at rest; most of 
the available energy would be used to produce the $\mu$ (or $\pi$) and
the kinetic energies of $\mu$ and $\nu_\mu$ (or $\pi$).

The total kinetic energy $Q_\mu$ in a decay
\begin{equation}
{\mbox {Pb}} \rightarrow \mu^\pm + \nu_\mu(\overline \nu_\mu)+
(A_1,Z_1)+(A_2,Z_2),
\label{eq:4}
\end{equation}
assuming close values of $A_1$ and $A_2$ (symmetric fission), is about 30 MeV 
for negative muons and about 20 MeV for positive muons. 
 Considering that the associated muon neutrino takes away a sizable fraction
of this energy, the spectrum of emitted muons could be like in a 
$\beta$ decay, with an average around $10 \div 15$ MeV. 

 The decay into two fragments would not lead to the best energetic situation
since the two resulting nuclei would on the average have large atomic mass and
 too large neutron numbers. 
Decays into more than two nuclear fragments would lead to more energetic 
muons; at the moment there are no estimates available for such decays.

There is only one experiment that published 
 upper limits for exotic decays of Pb \cite{canada}.
The  experiment used  123 g of natural lead, obtaining a $90\%$ C.L. upper
limit on the $\pi^0 \rightarrow \gamma \gamma $ counting rate of 
$3.3 \cdot 10^{-27}$ s$^{-1}$, for a total counting time of 109 hours.
The same authors searched also for pionic emission from
Uranium, and reported a $90\%$ C.L. upper limit for $\Gamma_\pi /
\Gamma_{SF} \sim 3.1 \cdot 10^{-4}$, where $\Gamma_{SF}$ is the width of the 
spontaneous decay fission. Using $^{235}$U and $^{252}$Cf 
they obtained upper limits of $1.4 \cdot 10^{-4}$ and $ 3.3 \cdot 10^{-10}$, 
 respectively.
A search for spontaneous emission of both muons and pions from $^{252}$Cf
 yielded upper limits for the ratio $\Gamma_{\mu, \pi} / \Gamma_{SF}$ in the 
range $10^{-6} \div 10^{-8}$ \cite{bucurescu}.

There are no experimental limits for muonic radioactivity.

In the following we describe a possible search for muonic radioactivity
of lead nuclei and we discuss the experimental set-up, the results of 
a Monte Carlo (MC) 
simulation giving estimates of the geometric efficiencies, a description of
possible backgrounds and the reachable limits.

\section{Experimental set-up}
\label{sec:setup}
We propose to perform an experimental search for muonic radioactivity from
lead nuclei in the low background conditions offered by the Gran Sasso 
underground Laboratory (LNGS). The very low cosmic muon flux and the 
low natural radioactivity of the rock in the experimental 
halls of the LNGS provide unique conditions, allowing a potential
discovery, or at least to establish a very good upper limit for
this exotic decay process.
 A detailed description of the different background sources is given in 
Sec. \ref{sec:background}.

\begin{figure}[t]
\begin{center}
       \mbox{  \epsfysize=7cm
               \epsffile{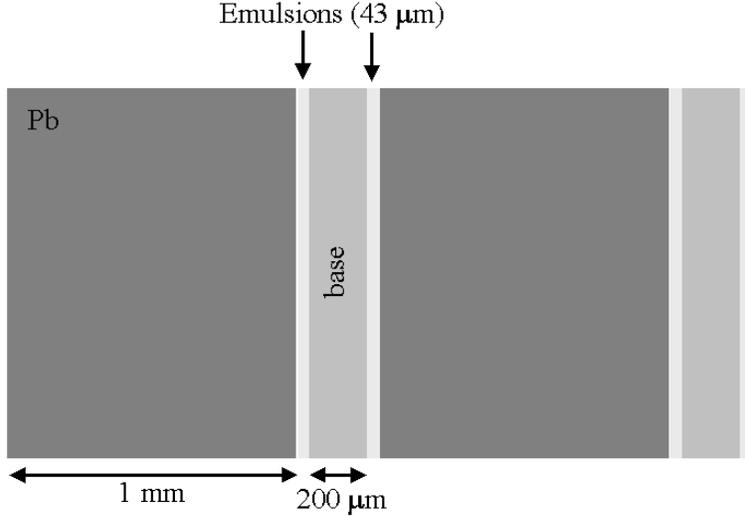}}
\caption{Illustration of the sequence lead/nuclear emulsion sheets which
characterise an OPERA brick.}
\label{fig:ECC}
\end{center}
\end{figure}

This search could be a by-product of the OPERA experiment \cite{opera} 
presently under construction at LNGS. OPERA is a hybrid long-baseline neutrino 
experiment, aimed to the direct observation of $\nu_\tau$ appearance 
from  $\nu_\mu \longleftrightarrow \nu_\tau$ oscillations in the 
CERN-Gran Sasso $\nu_\mu$ beam. The detector is made of 2 supermodules, each 
with a massive lead/nuclear emulsion target, electronic detectors and 
a magnetic spectrometer. Nuclear emulsions are used 
as high resolution ($0.06~\mu$m) 
tracking devices for the direct detection of the decay of the 
$\tau$ leptons produced in the charged current $\nu_\tau$ interactions with 
the target. 

\begin{figure}[ht]
\begin{center}
       \mbox{  \epsfysize=8cm
               \epsffile{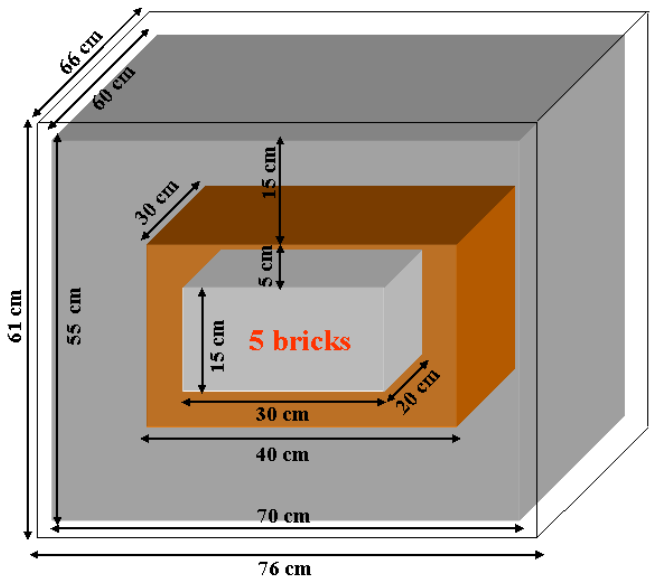}}
\caption{Sketch of the shield used for background rejection/reduction. It 
is composed by layers of very pure copper and lead, plus an outside layer 
of polyethylene (``transparent'' box). The set-up is located in
the OPERA emulsion storage room, equipped with a ventilation system.}
\label{fig:schermo}
\end{center}
\end{figure}

The OPERA base element (``brick'') is composed of 56 lead sheets (1 mm thick) 
interleaved with 2 nuclear emulsion sheets (43 $\mu$m thick) on both sides 
of a 200 $\mu$m thick plastic base. This sequence is illustrated in Fig. 
\ref{fig:ECC}. The area of each sheet is $10.3 \times 12.8$ cm$^2$.

The OPERA lead bricks (each containing a mass of 8.23 kg of Pb) could 
allow an experimental search for 
muon emission from lead, with exposures of several months. Their
analyses with the fast automatic optical microscopes \cite{microscopi} 
would establish the local background contributions and validate the 
analysis procedures.

\begin{figure}[htb]
\vspace{-12mm}
\begin{center}
       \mbox{  \epsfysize=8cm
               \epsffile{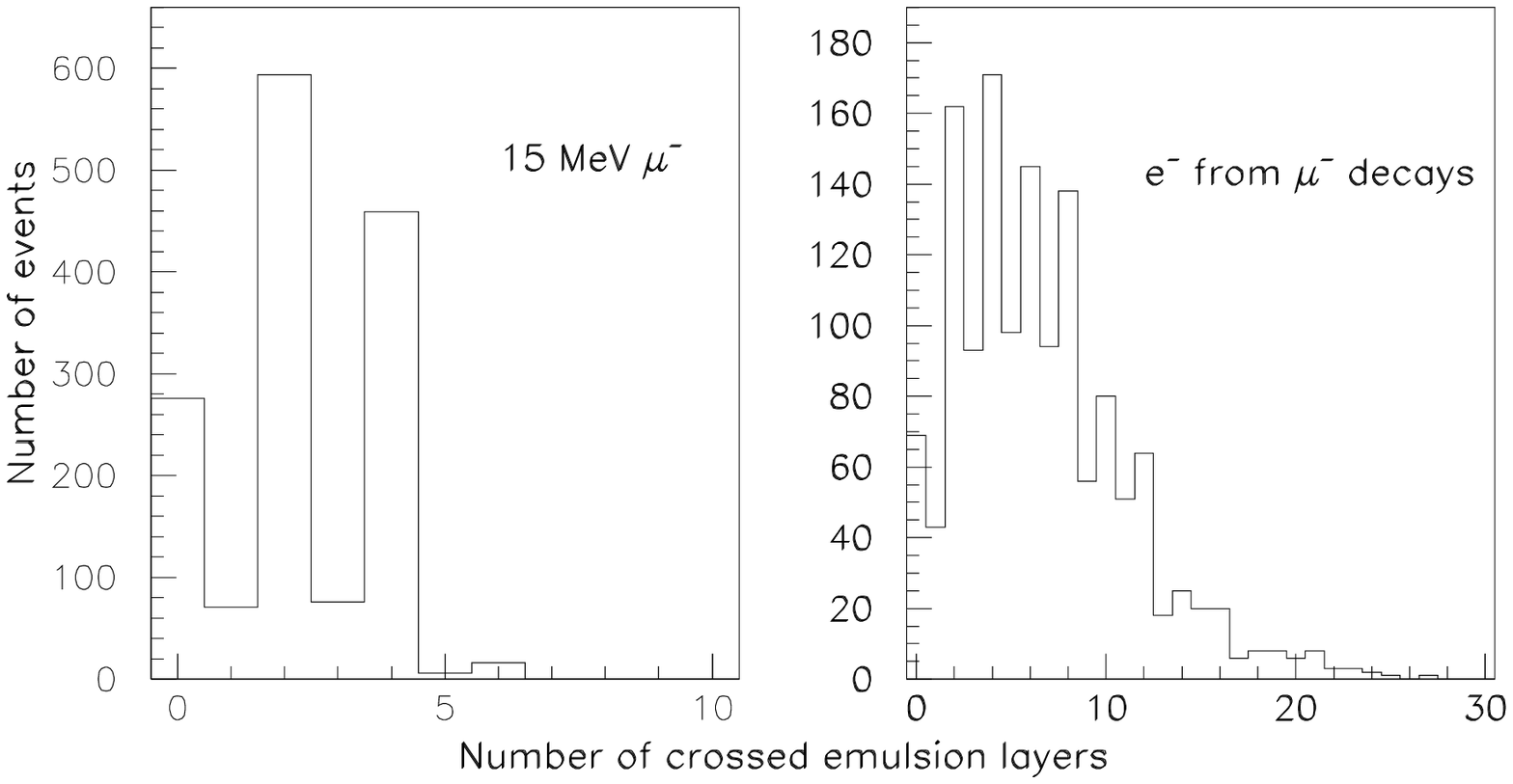}}
\caption{Distribution of the number of
emulsion layers crossed by the simulated $\mu^-$ (left) and by the electron
resulting from its decay (right), assuming an initial muon energy of 15 MeV
and isotropically distributed emission angles.}
\label{fig:mu-e-15mev}
\end{center}
\end{figure}

A preliminary test using a ``small'' brick composed of 22 lead sheets and 21
recently refreshed nuclear emulsion sheets was started at the end of March
2005. As the background rejection/reduction (see Sect. \ref{sec:background}) 
is a crucial point for this search, the detectors are surrounded on 
all sides by a shield, making a closed box structure similar to those 
used in experiments for dark matter and for neutrinoless double beta 
decay searches. The shield is composed of an inner layer 5 cm thick of 
very pure copper followed by 15 cm of very low 
activity lead\footnote{These thicknesses have been found to be adequate 
with 
a Monte Carlo simulation considering the effects of $0.5 \div 2.6$ 
MeV photons, which  could produce electrons mimicking the searched events.}. 
 The third and last layer of the shield is a 3 cm thick polyethylene, in order 
to 
absorb neutrons. A scheme of the installed shield, with the dimensions for
containing 5 OPERA bricks, is presented in Fig. \ref{fig:schermo}.
 The set-up is located in the emulsion storage room, in hall B of the Gran 
Sasso Laboratory. The radon reduction is obtained with a ventilation 
system due to fresh air forced circulation.

The mean range of 15 MeV muons in lead is about 1.5 mm, so most of
the possibly produced muons could be detected in the two layers of emulsion
 close to the Pb sheet ``source''; depending on the angle of emission and
on the location of the decay point, some of 
them could also be detected by the next emulsion sheets. 

It is clear that
thinner lead sheets would allow the emitted muons to loose a lower fraction
of their energy to escape from the lead and to cross more emulsion
layers. They would also yield a higher detection efficiency.

\section{Monte Carlo simulation}
\label{sec:MC}

A MC simulation program was implemented to estimate the occurrence of 
different event topologies. The simulation is based on the GEANT3 \cite{geant}
 package applied to the OPERA lead/emulsion set-up. At this point 
it does not include the complete detector response and does not 
estimate the real event reconstruction efficiencies.

The simulation reproduces one complete OPERA brick, where muons of different 
energies (see the first column of Table \ref{tab:mu}) 
originate in random positions in the lead sheets. The initial muon 
directions are isotropically generated. We assumed different definitions 
for a candidate event, requiring that the muon crosses at least: $(i)$ one 
single emulsion layer, $(ii)$ two emulsion layers (near the same base), and
$(iii)$ three emulsion layers (and thus also a lead plate, which would allow 
a better identification using the measured energy loss). We also requested the 
detection of the decay positron or electron in at least $(iv)$ two, $(v)$ 
three, or $(vi)$ five emulsion layers, together with the muon detection.

\begin{figure}[ht]
\begin{center}
       \mbox{\hspace{-0.8cm}
             \epsfysize=4.5cm
             \epsffile{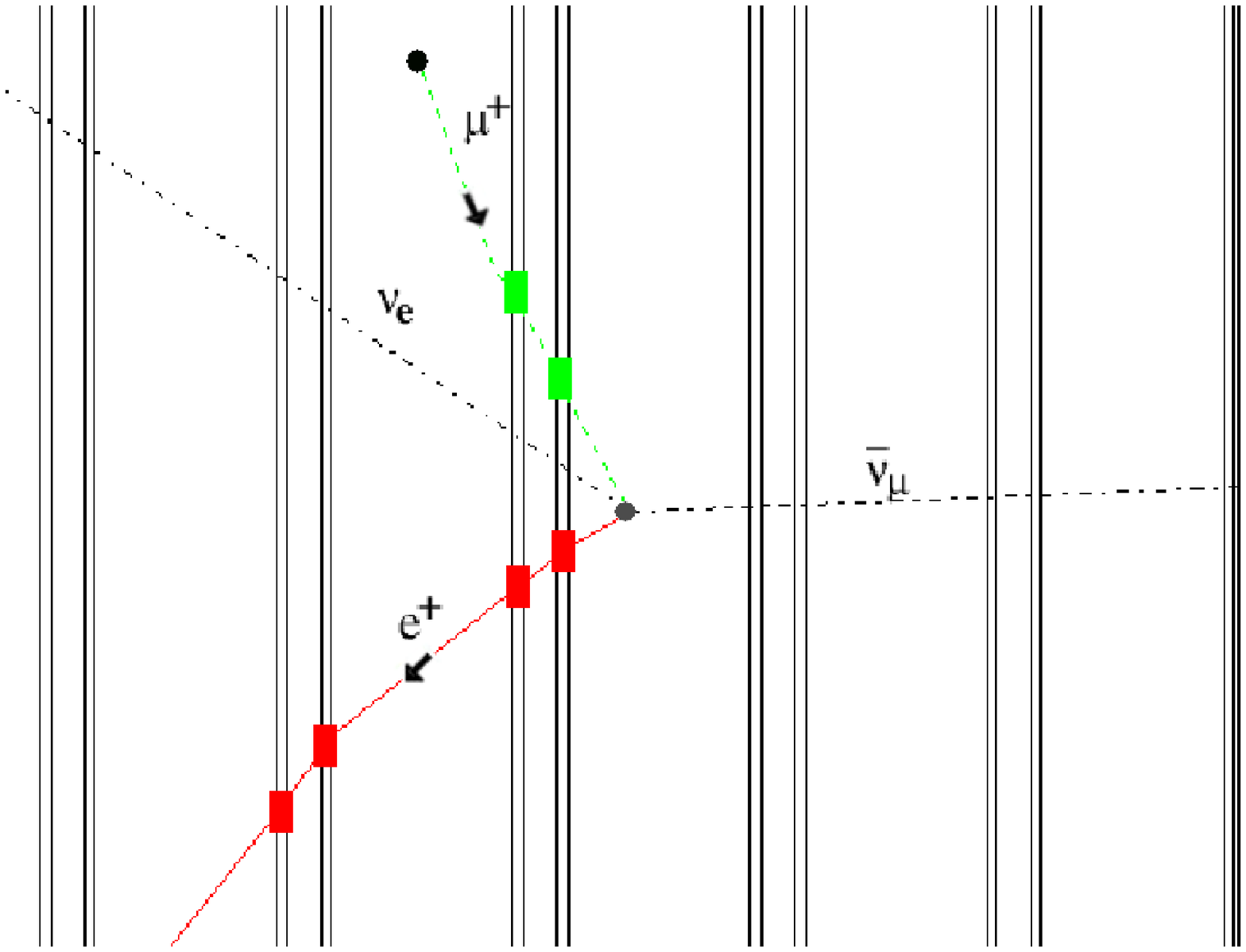}
               \hspace{0.7cm}
               \epsfysize=5cm
               \epsffile{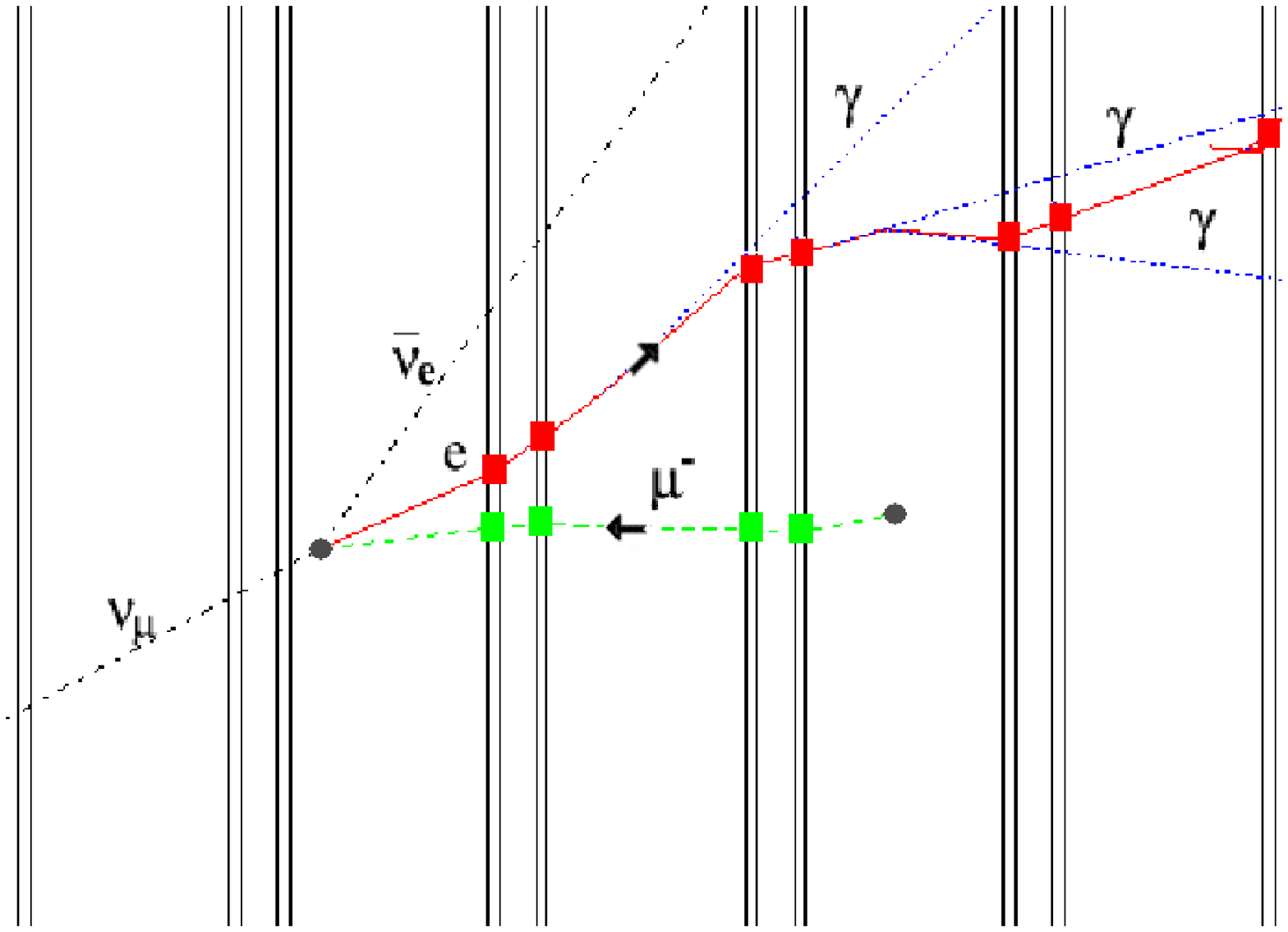}}
(a) \hspace{6cm} (b)
\caption{Simulated events of spontaneous $\mu^+$ (a) and $\mu^-$ (b)
emission from lead inside an OPERA brick, assuming an initial muon kinetic 
energy of 
15 MeV. In (a) the $\mu^+$ starts from the second lead sheet and stops in the  
 third one, where it decays into $e^+ \nu_e \overline \nu_\mu$. 
The $\mu^+$ is seen in 2 emulsion sheets, the $e^+$ in 4. 
In (b) the $\mu^-$ starts from the fourth lead sheet and stops in the  
second one, where it decays into $e^- \overline \nu_e \nu_\mu$. 
The $\mu^-$ is seen in 4 emulsion sheets, the $e^-$ in $>6$.
The change of direction for the $\mu^\pm$ and $e^\pm$ in the interfaces 
lead-emulsion are due to the MC approximation which, for each material, adds 
the multiple Coulomb scattering effects and gives a global scattering angle 
of the outgoing direction relative to the incoming one.    
}
\label{fig:mu}
\end{center}
\end{figure}

Fig. \ref{fig:mu-e-15mev} shows the distribution of the number of
emulsion layers crossed by a sample of simulated $\mu^-$ (left) and by 
the electrons
resulting from their decay (right), assuming an initial muon energy of 15 MeV 
and isotropically distributed emission angles.

{\small
\begin{table}[htb]
\begin{center}
\begin{tabular}{|c|c|c|c|c|c|c|c|c|c|c|c|c|}
\hline
$N_{\mu}$ & 2 & 2 & 2 & 2 & 4 & 4 & {\bf 4} & 4 & 6 & 6 & 6 & 6 \\
$N_{e}$   & 0 & 2 & 4 & 6 & 0 & 2 & {\bf 4} & 6 & 0 & 2 & 4 & 6 \\
\hline
5 MeV  & 12 & 10 & 9  & 7  & - & - & {\bf -} & - & - & - & - & - \\
7 MeV  & 25 & 23 & 21 & 14 & - & - & {\bf -} & - & - & - & - & -  \\
10 MeV & 51 & 46 & 41 & 31 & 0.6 & 0.5 & {\bf 0.4} & 0.3 & - & - & - & - \\
15 MeV & 74 & 66 & 58 & 44 & 31 & 28 & {\bf 23} & 17 & 1 & 1 & 0.9 & 0.7  \\
20 MeV & 84 & 72 & 63 & 49 & 56 & 50 & {\bf 40} & 30 & 29 & 26 & 21 & 15  \\
25 MeV & 89 & 75 & 66 & 50 & 70 & 60 & {\bf 49} & 36 & 48 & 42 & 34 & 25 \\
30 MeV & 92 & 75 & 67 & 52 & 78 & 66 & {\bf 54} & 36 & 60 & 51 & 42 & 28 \\
\hline
\end{tabular}
\caption {Percentage of events as a function 
of the minimum number of emulsion films crossed by both $\mu$ $(N_\mu)$ 
and $e$ $(N_e)$, for different $\mu$ initial energies (col. 1). The 
column in bold refers to the conditions requested by the OPERA 
tracking system to define a track $(N_{\mu}=N_{e}=4)$. }
\label{tab:mu}
\end{center}
\end{table}
}

 As there is no theoretical prediction concerning the muon 
energy spectrum, an overall fraction of events with a defined topology
cannot be estimated. The percentages of events listed in 
Table \ref{tab:mu} were computed for samples of MC events 
with fixed energies. The detection threshold for the $\mu^\pm$, obtained 
requiring at least 2 nuclear films crossed by the $\mu^\pm$, is about 
3 MeV. For each topology, the values quoted 
in Table \ref{tab:mu} were obtained as averages over all
 possible muon emission directions. These estimates may be considered 
as geometrical efficiencies $\epsilon_g$. 

In Fig. \ref{fig:mu} are shown two simulated 
muon emission events from the lead inside an OPERA brick, assuming an 
initial kinetic energy of 15 MeV, and the production of a $\mu^+$ in Fig. 
\ref{fig:mu}a and a $\mu^-$ in Fig. \ref{fig:mu}b. The dashed lines (longer 
lines) represent the muon trajectories, the solid lines are positrons or
electrons, the dash-dotted lines are neutrinos and the dotted lines
are photons. In this two favourable cases the muon and the electron/positron 
would be easily identified.

\subsection{Estimates of global detection efficiencies}
\label{sub:eff}
 The geometrical $\epsilon_g$ have to be multiplied by the $\mu^\pm$ and 
$e^\pm$ reconstruction efficiencies to obtain the total efficiency. We 
made a first estimate of $\epsilon_{tot}$ for muons 
based on the measurement procedures  and the algorithms used in OPERA 
\cite{microscopi}. 

The relation between the energy of charged particles and the density of grains
produced in emulsions along their trajectories \cite{atlante-emulsioni} shows 
that muons with the predicted energies would give recognisable black 
tracks without 
relevant scattering in the emulsion films. For the decay electrons/positrons 
the tracks would be similar, but with a lower density of grains 
per unit path-length. 

With the OPERA tracking procedure \cite{saverio}, the mean detection 
efficiency for each ``microtrack'' in one emulsion 
film is $\simeq 95\%$; it is a value averaged on the angular range 
between tracks normally incident on the plate surface and tracks with 
incident angles $\simeq 0.8$ rad with respect to the normal. The 
instrumental limit of 0.8 rad on the incident 
direction of each microtrack introduces an event selection of $\sim 30\%$.

The ``base track'' (obtained from 2 microtracks separated by 
the plastic base) reconstruction efficiency of 
$\simeq 90\%~(95\% \times 95\%)$ is comprehensive of the uncertainties 
due to the emulsion shrinkage and possible surface distortions. The 
top-bottom linking efficiency ranges from the $\sim 50\%$ 
for 5 MeV particles to $\sim 99\%$ for particles with energies $\geq 20$ MeV.
It is mainly due to the multiple scattering in the plastic base.
 The base track linking efficiency is $\sim 6\%$ for the whole range 
of initial muon energies.  

In order to reduce the  
background, an algorithm defines a ``track'' requiring at least 4 
emulsion films (it means at least 1 lead sheet and 2 plastic bases) 
crossed by a particle. The detection efficiency $\epsilon_{tot}$ 
for a muon crossing at least 4 emulsion 
films is the product of the percentage of events with these topological 
requirements $\epsilon_g$ (col. 6 of Table \ref{tab:mu}), the 30\% 
given by the angular limit, 4 times the 95\% 
efficiency for the microtracks, 2 times the top-bottom linking efficiency and
the linking efficiency between 2 base tracks. As shown in Table 
\ref{tab:mu}, the threshold muon 
energy to have at least 4 emulsion films crossed by the muon is 10 MeV.
         
For the electrons, the probability of crossing at least 4 emulsion films 
$\epsilon_g^e$ is about 68\% for the whole range of initial muon energies. 
 This percentage was obtained from the simulated events requiring a minimum  
number of 4 emulsion films crossed by the electron and no constraints on the
muon trajectory. The convolution 
with the topological requirements of the 
angular directions within 0.8 rad (30\% selection), the 95\% efficiency 
reconstruction of the 4 microtracks, the top-bottom linking efficiency 
and the linking efficiency between base tracks yields the total electron 
detection efficiency.

{\small
\begin{table}[htb]
\begin{center}
\begin{tabular}{|c||c|c||c|c|c|}
\hline
&\multicolumn{2}{c||}{$N_\mu=2$}&\multicolumn{3}{c|}{$N_\mu=4$} \\
\hline
$E_\mu$ & $\epsilon_g$ & $\epsilon_{tot}^{OPERA}$ & 
$\epsilon_g$ & $\epsilon_{tot}^{OPERA}$ & $\epsilon_{tot}^*$  \\
\hline
5 MeV  & 12  & 1.5   &  -  & -        & - \\
7 MeV  & 25  & 4.3   &  -  & -        & - \\
10 MeV & 51  & 12    & 0.6 & 0.006    & 0.1 \\
15 MeV & 74  & 19    & 31  & 0.4      & 5.6 \\
20 MeV & 84  & 22    & 56  & 0.8      & 11  \\
25 MeV & 89  & 24    & 70  & 1        & 14  \\
30 MeV & 92  & 25    & 78  & 1.1      & 16  \\
\hline
\end{tabular}
\caption {Geometric efficiencies $(\epsilon_g)$ and total muon detection 
efficiencies $(\epsilon_{tot}^{OPERA})$ as a function of  
the initial muon energy computed on the basis of the OPERA
tracking procedure for $N_\mu=2$ (col. 2-3) and for  
$N_\mu=4$ (col. 4-5). The values $\epsilon_{tot}^*$ (col. 6)
are obtained relaxing the angular and position tolerances in the 
base track reconstruction (see text). All values are in \%.}  
\label{tab:eff}
\end{center}
\end{table}
}

We intend to optimise the tracking procedure for this search using new
scanning conditions. It may be possible to require 
only 2 emulsion films crossed by a muon to reconstruct 
its track. This optimisation should be possible, because of the easily 
recognisable black tracks produced by slow muons. Once a muon is found, we
should be able to check visually the presence of the decay electron.

Table \ref{tab:eff} shows the \% efficiencies 
$\epsilon_g$ and $\epsilon_{tot}^{OPERA}$ computed requiring
at least 2 emulsion films crossed by the muon (col. 2-3) and 4   
emulsion films crossed by the muon (col. 4-5). The values refer to the present
OPERA tracking \cite{saverio}. The 
linking efficiencies may be improved
relaxing the angular and position tolerances between consecutive base tracks.
This should be possible because of the low background conditions.
Requesting an angular tolerance of 100 mrad and a position tolerance of
100 $\mu$m, we obtained the values given in col. 6 of Table \ref{tab:eff}. 

In OPERA the different emulsion sheets of each brick are aligned mechanically
and then exposed to cosmic ray muons to have 
a relative alignment of $\sim 1~\mu$m \cite{cosmici}. In the present 
case, due to the low 
background conditions, we expect that the mechanical alignment precision 
should be adequate for the purpose; since the emulsion layers are
placed horizontally, a check of the alignment could come from the few high 
energy 
cosmic ray muons which should cross the emulsions \cite{trd-macro}.

As the expected halflifes $t_{1/2}$ are much larger 
than any reasonable exposure time $T$, the expected sensitivities are 
estimated from
\begin{equation}
\frac{\delta N}{N_0} = \frac{\ln 2}{t_{1/2}}T \epsilon_{tot}
\label{eq:6}
\end{equation}
where $\delta N = 2.3$ is the number of events corresponding to a 90\% C.L. 
limit assuming no candidates, $N_0$ is the initial number of nuclei, and 
$\epsilon_{tot}$ is the experimental efficiency. 

Assuming the use of one OPERA brick for one year exposure and a global 
detection efficiency of $\sim 10\%$, we could reach a 
sensitivity of about $7 \cdot 10^{23}$ yr ($90\%$ C.L.).

\section{Background estimates}
\label{sec:background}

The background originates from several different sources.

1) The environmental radon background is reduced by the ventilation  
 and may be monitored using nuclear track detectors, such 
as CR39 or Makrofol, that are insensitive to muons.

2) The background produced by the ambient neutron flux ($\sim 0.42 
\cdot 10^{-6}$ neutrons/s/cm$^2$, with energy greater than  
1 MeV \cite{arneodo}) should be studied; its effect can be reduced 
by a proper shielding and appropriate ``event definitions".

3) The background produced by the radioactive $^{210}$Pb isotopes present in 
the lead. The $\alpha$ and $\beta$ radioactivity of the lead used in OPERA 
\cite{opera} is $< 0.02$ Bq/g. A single lead plate has an 
activity $< 3$ Bq, and only a fraction of the particles emitted from
the outer surfaces of the material escapes from the lead sheets. The
 $\alpha$ background could be eliminated by thin plastic foils inserted
 between the lead plates and the emulsion sheets.

4) The background produced by radioactive nuclides present in the 
emulsion, for example $^{212}$Po nuclei. The emitted $\alpha$ particles have 
an energy of 8.785 MeV and a range in emulsion of 74 $\mu$m. This background
should be reducible by dE/dx measurements on the tracks and from range 
considerations.

5) The background due to cosmic rays.

$(i)$ Cosmic ray (CR) muons reaching from above the LNGS
underground labs (3700 hg/cm$^2$  average rock cover) are $\sim 1$ 
muon/h/m$^2$ \cite{trd-macro}; the
number of CR muons expected to cross one brick during one year is about
90, sufficient to improve alignments of the different emulsion
layers. They have an average energy of 240 
GeV \cite{trd-macro}. Tracks produced by CR muons in emulsion may be 
easily identified
 on the basis of their high energy and relatively small energy loss. 
 Low energy CR muons can be removed
 requesting that the tracks start in a lead plate.

$(ii)$ Pions produced by muons interacting in the lead plates. Slow pions 
decay into muons and simulate a spontaneous muon emission. 
This process was investigated by MC techniques in MACRO, since it was a 
background for the measurement of the upgoing 
muon flux \cite{fondo}; the probability to produce a pion by a CR muon 
through photonuclear interactions in a layer of 150 g/cm$^2$ was estimated 
to be  $<2 \cdot 10^{-5}$, and the fraction of low energy backward 
scattered pions is $<2 \cdot 10^{-4}$. Since in an OPERA brick the yearly 
rate of charged pions produced inside the plates is  
$<6 \cdot 10^{-4}$, this background may be neglected.

$(iii)$  The flux of muons resulting from atmospheric $\nu_\mu$ interactions 
around the detector is known from measurements done by MACRO 
\cite{macro}; they yield a negligible background.

$(iv)$ In the Gran Sasso underground laboratories 
 the CR neutron flux is about $\sim 10$ neutrons per square centimetre in one 
year \cite{arneodo,n1}. Only a fraction of these neutrons interact in the 
detector and the events produced by them should be recognisable in emulsions.

6) The future CERN-Gran Sasso neutrino beam may yield a background 
contribution; the energies of muons induced by the $\nu$ beam
will be much higher than the energies of emitted muons from lead. Moreover, 
 the beam will be continuously monitored. Also 
this background should be easily recognised and removed. The beam
will not be present before the middle of 2006.

\section{Conclusions and perspectives}
\label{sec:conclu}

We described a test search for spontaneous emission of muons from 
Pb nuclei, using some OPERA lead/emulsion bricks. The proposed test is 
suggested by some theoretical works performed in the last years; it 
would be complementary to other experiments looking for such 
exotic radioactivities from heavier nuclei and could reach a good 
sensitivity level. For a visual fine-grained technique (the emulsion 
technique) the identification is easier and the background lower. The 
test will yield the first information relative to the background in the 
OPERA bricks.

We are in the process of making a complete simulation of the detector 
including its response and the track reconstruction efficiencies. As
already stated, thinner lead sheets would 
improve the scanning and the track reconstruction efficiencies.

 We have shown that stringent limits for spontaneous muon radioactivity
may be reached. We would obtain $t_{1/2} \geq  7 \cdot 10^{23}$ years.

In the future, one could repeat the experiment using elements heavier 
than lead.

\section{Acknowledgements}

We thank many colleagues for encouragements and discussions, in particular 
G. Bonsignori. We acknowledge the cooperation of E. Bottazzi, C. Bucci, A. 
Candela, A. Corsi, L. Degli Esposti, M. Laubenstein, E. Tatananni, V. Togo 
and C. Valieri.

\end{document}